\documentclass[aps,prl,preprint,superscriptaddress]{revtex4}
\usepackage{amsfonts}

\usepackage{graphicx}

\def\Blue{} % PANTONE BLUE-072
\def\Black{} % PANTONE PROCESS-BLACK
\def\Red{} % PANTONE RED

\begin{document}

\title{ Entwined Paths, Difference Equations and the Dirac Equation}
\author{G.N. Ord}
\email[ corresponding author ]{gord@acs.ryerson.ca}
\affiliation{M.P.C.S. \\
Ryerson University\\
Toronto Ont.}
\author{R.B. Mann}
\email{mann@avatar.uwaterloo.c}
\affiliation{Dept of Physics\\
University of Waterloo\\
Waterloo, Ont Canada}
\date{May 18 2002}

\begin{abstract}
Entwined space-time paths are bound pairs of trajectories which are
traversed in opposite directions with respect to macroscopic time. In this
paper we show that ensembles of entwined paths on a discrete space-time
lattice are simply described by coupled difference equations which are
discrete versions of the Dirac equation. There is no analytic continuation,
explicit or forced, involved in this description. The entwined paths are
`self-quantizing'. We also show that simple classical stochastic processes
that generate the difference equations as ensemble averages are stable
numerically and converge at a rate governed by the details of the stochastic
process. This result establishes the Dirac equation in one dimension as a
phenomenological equation describing an underlying classical stochastic
process in the same sense that the Diffusion and Telegraph equations are
phenomenological descriptions of stochastic processes.
\end{abstract}

\date{June 8 2002}
\maketitle

% number in the upper right hand corner of the title page in preprint mode.
%\preprint{Ryerson University}


%\thanks{}

%\thanks{}

\section{Introduction}

This paper is the first of a series of papers exploring the consequences of
a recent discovery. The new result is that the Feynman Chessboard model of
the Dirac propagator can be transplanted into classical statistical
mechanics, bypassing Formal Analytic Continuation(FAC) completely\cite%
{gord01a}. The new feature allowing this is the use of entwined space-time
paths which are essentially self-quantizing. These will be discussed
shortly, but first we clarify what is meant by the phrase `transplanted into
classical statistical mechanics'.

One usually arrives at quantum mechanics through one of two routes. The most
common approach is to analytically continue from $\mathbb{R}$ to $\mathbb{C}$
explicitly by imposing operator relations (eg. $p\rightarrow -i\hbar \frac{%
\partial }{\partial x},E\rightarrow i\hbar \frac{\partial }{\partial t}$). \
Another alternative is to force FAC by imposing physical conditions that
cannot be met within the original space (eg. requiring diffusion to be
reversible).\cite{Nelson66,Nelson85,Nagasawa96,Nottale92,Naschie95c} Both of
these procedures extend classical physics to a suitably enlarged regime
(i.e. Hilbert space). However the resultant wavefunctions are formal objects
with no direct interpretation.

We avoid these routes entirely. In our approach the components of
wavefunctions are classical ensemble averages obtainable by simple counting
processes. It is the spatio-temporal geometry of entwined paths that gives
rise to the properties described by the standard quantum mechanical complex
wave equations. The relevant algebraic property $i^{2}=-1$ appears
explicitly through the geometry of entwined paths.
% Counting paths with
%entwined geometry requires the use of an antihermitean matrix which
%provides the link with the relevant algebra.
 It is because the equivalent
of this algebraic structure (expressed, as we shall later
see, by an antihermitian operator) is built into the geometry of the
space-time paths themselves that we never need to introduce it
artificially through a FAC. This is the main feature of our results and we
will return to the relation between geometry and algebra in the discussion
at the end of the paper.

By avoiding the FAC generally used to quantize a system, entwined paths
provide the Dirac equation with a new context that is conceptually very
different from its context in quantum mechanics. In quantum theory the Dirac
equation is a `fundamental' equation. Wavefunction solutions are thought to
contain `all the information about the state of a system', and the Dirac
equation describes the evolution of the wavefunction.

In the context described in this paper, the Dirac equation is not a
fundamental equation at all. Rather it is a phenomenological equation that
describes ensemble averages of a classical charge density arising from
entwined paths. Just as the Diffusion equation is a phenomenological
equation describing a density of random paths (Brownian motion), so the
Dirac equation describes net densities of entwined paths, where the
time-reversed portions of paths add the new qualitative features of
interference and reversibility. Because there is a specific underlying
stochastic model involved, wavefunction solutions do not contain all the
information about the state of a system, they are simply ensemble averages
of the background stochastic process. In the new context, the stochastic
process itself is the fundamental object. This means that there is an
identifiable underlying stochastic process involved in the \emph{formation}
of wavefunctions.

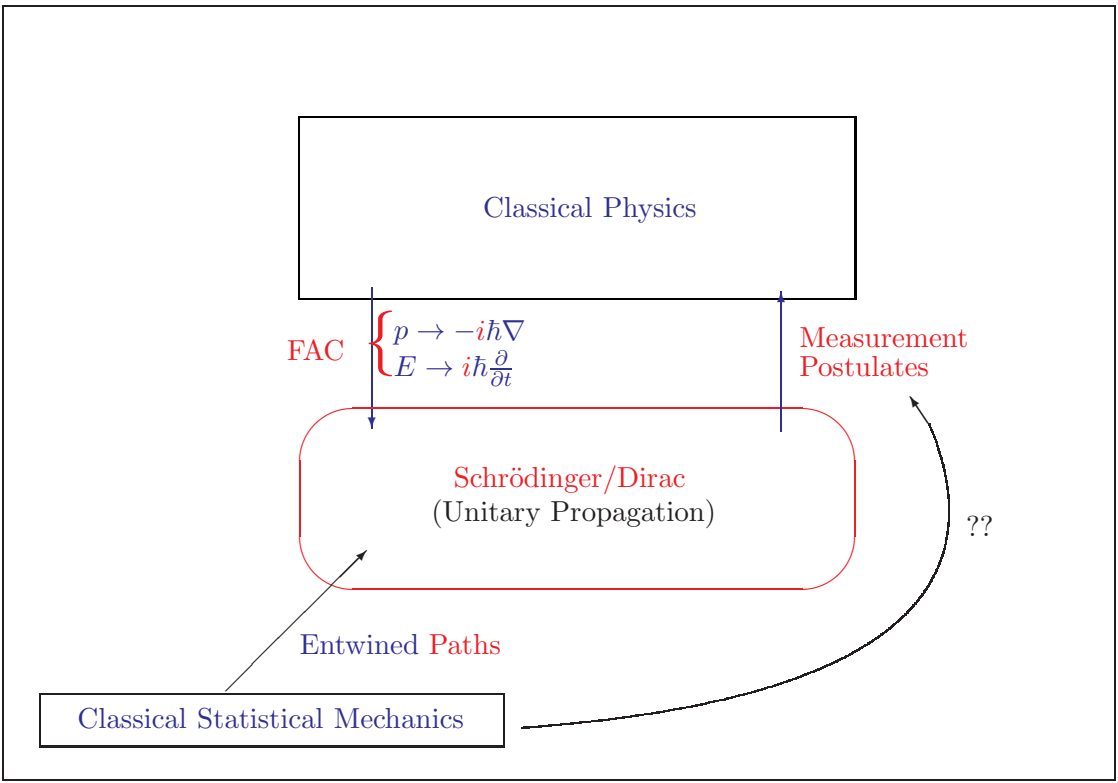
\begin{figure}[tbp]
\vspace{2.5cm} \hspace{-6cm} 
\begin{picture}(7,20)(-00,-40)
\setlength{\unitlength}{14pt}
\put(-2,-.4){\framebox(15,4.9){}}
\Blue
\put(2.7,1.8){ Classical Physics}
\put(-.05,-.1){\vector(0,-1){3.8}}
\put(-.5,-1.5){ $\quad p\to - \Red i \Blue \hbar \nabla$}
\put(-.5,-2.5){ $\quad E\to \Red i \Blue \hbar
\frac{\partial}{\partial t}$}
\put(-2.6,-2.05){{\Red{ FAC  }  \huge \{ }}
\put(5.5,-5.8){\oval(15,4.9){}}
\put(1.9,-5.5){ Schr\"{o}dinger/Dirac }
\Black
\put(1.3,-6.4){ (Unitary Propagation) }
\Blue
\put(11,-4){\vector(0,1){3.8}}
\Red
\put(11.5,-1.7){{Measurement }}
\put(11.5,-2.5){{Postulates }}
\Black
\put(-4,-11){\vector(1,1){3.8}}
\put(-2,-10){{\Blue Entwined \Red  Paths }}
\put(-8,-12){{\Blue Classical Statistical Mechanics }}
\Black
%\put(10,-7.5){\line(1,-1){3.8}}
%\put(13,-12){{\Blue Born ??? }}
\Black
\put(-10,-13.4){\framebox(30,20.9){}}
\put(-9.,-12.5){\framebox(12.5,1.4){}}
%\put(4,-12.){\line(1,0){3.8}}
 \qbezier(4,-12.)(18,-11)(15,-3.8)%origin to asymptote
\put(15,-3.8){\vector(-2,3){.5}}
\put(16,-6.8) {??}
\end{picture}
\vspace{5cm}
\caption{In conventional interpretations of quantum mechanics we pass from
classical physics (upper box) to the Schr\"{o}dinger or Dirac equation
(rounded box) through a Formal Analytic Continuation (FAC). This brings
wave features into the classical particle paradigm as wavefunctions
propagate unitarily. Measurement postulates are then used to interpret
wavefunction solutions in terms of macroscopic measurements. This paper
discusses an alternative route to the Dirac equation from classical
statistical mechanics using `entwined paths'(lower left). This route does
not require a FAC since the geometry of the paths automatically build in
the relevant wave behaviour. Furthermore, in the new context, the Dirac
equation appears as a phenomenology for the underlying stochastic process.
Further examination of the stochastic process is required to see if the
measurement postulates can also be supported in the new context.}
\label{fig1}
\end{figure}

 In contrast, quantum mechanics has nothing to say about the process of
wavefunction formation, since there, the wavefunction is just
part of an algorithm. Probability only enters quantum mechanics
through the measurement postulates, not through unitary evolution
(Fig(\ref{fig1})). Thus, in quantum mechanics we postulate that the
modulus squared of the wavefunction represents a probability density. That
the postulate is correct is well verified experimentally, but remains a
feature which does not follow from unitary evolution itself. On the other
hand entwined paths support unitary evolution of an ensemble average which
itself has an underlying
 stochastic process that, as we shall see,  is amenable to  direct
simulation. In the future we shall be able to ask `does the underlying
stochastic process also mimic the measurement postulates through the
stochastic formation of the wavefunction?' If the answer is yes, then the
stochastic process we are proposing for the Dirac equation may have a
deeper connection to quantum mechanics. If not the two contexts for the
Dirac equation will remain distinct. 

\bigskip

\section{Entwined Paths}

We shall be working in a two dimensional discrete space $(z,t)$ with lattice
spacings $\delta$ and $\epsilon$ respectively. Although we shall eventually
think of $t$ as time, it is convenient at this point to think of $t$ as a
spatial coordinate. Entwined paths can be generated in two different ways,
both ways being important in understanding the resulting phenomenology. We
shall employ two walkers, Eve and Max, who will generate and colour random
walks, using the two different methods. Eve's method is explicitly Ergodic
in the sense that  she will generate the full ensemble of entwined paths in
a serial fashion after a sufficiently long time. Max will not use an
explicitly ergodic process. He will however generate an ensemble of paths
using a Markovian process which generates paths in parallel, many at a
time. At each step, Max's behaviour will depend only on his current state.
Both walkers generate and colour their paths through
$(z,t)$ based on calls to a random process R which generates a string of
letters, with the distribution:

\begin{equation}
R=\left\{ 
\begin{array}{lr}
U & \mbox {with probability }1-a\Delta t \\ 
M & \mbox {with probability }\quad a\Delta t.%
\end{array}%
\right.  \label{epsilons}
\end{equation}%
We can think of $M$ as standing for Marked and $U$ standing for Unmarked,
with the whole string of symbols a sequential list or tape $T$.

Eve's instructions for colouring her path through the lattice are as
follows. Eve starts at the origin and her first step is to the lattice site $%
(\delta ,\epsilon )$. At the next step she consults the random process. If a 
$U$ is the first letter on the tape she maintains her direction and steps to 
$(2\delta ,2\epsilon )$. If she gets an $M$ she changes her direction in $z$
and steps to $(0,2\epsilon )$. Again she consults the tape. If she gets a $U$
she maintains her direction and takes another step. If she gets her second $M
$ she drops a `marker' at the site but steps forward maintaining her
direction. This process is repeated, alternately changing direction and
dropping markers whenever the random process indicates a mark. Eve colours
her path blue as she goes. With each step she advances one lattice spacing
in $t$ until the first time she is called to drop a marker after a set
return `time' $t_{R}$. This marker will be the last symbol on the tape that
describes this particular path. At this marker Eve maintains her direction
in $z$ but steps back one unit in $t$ and switches to the colour red (Fig
2.A). Eve now makes her way back to the origin down the `light-cone' paths
(paths with slopes of unit magnitude in the $(z,t)$ plane) which intersect
all of the markers, colouring the path red as she goes (Fig(2.B)). On her
return path Eve doesn't need to consult a tape since her trail of markers
uniquely defines her return path.

%%%%%%%%%%%%%%%
\begin{figure}[tbp]
\includegraphics[scale = 1]{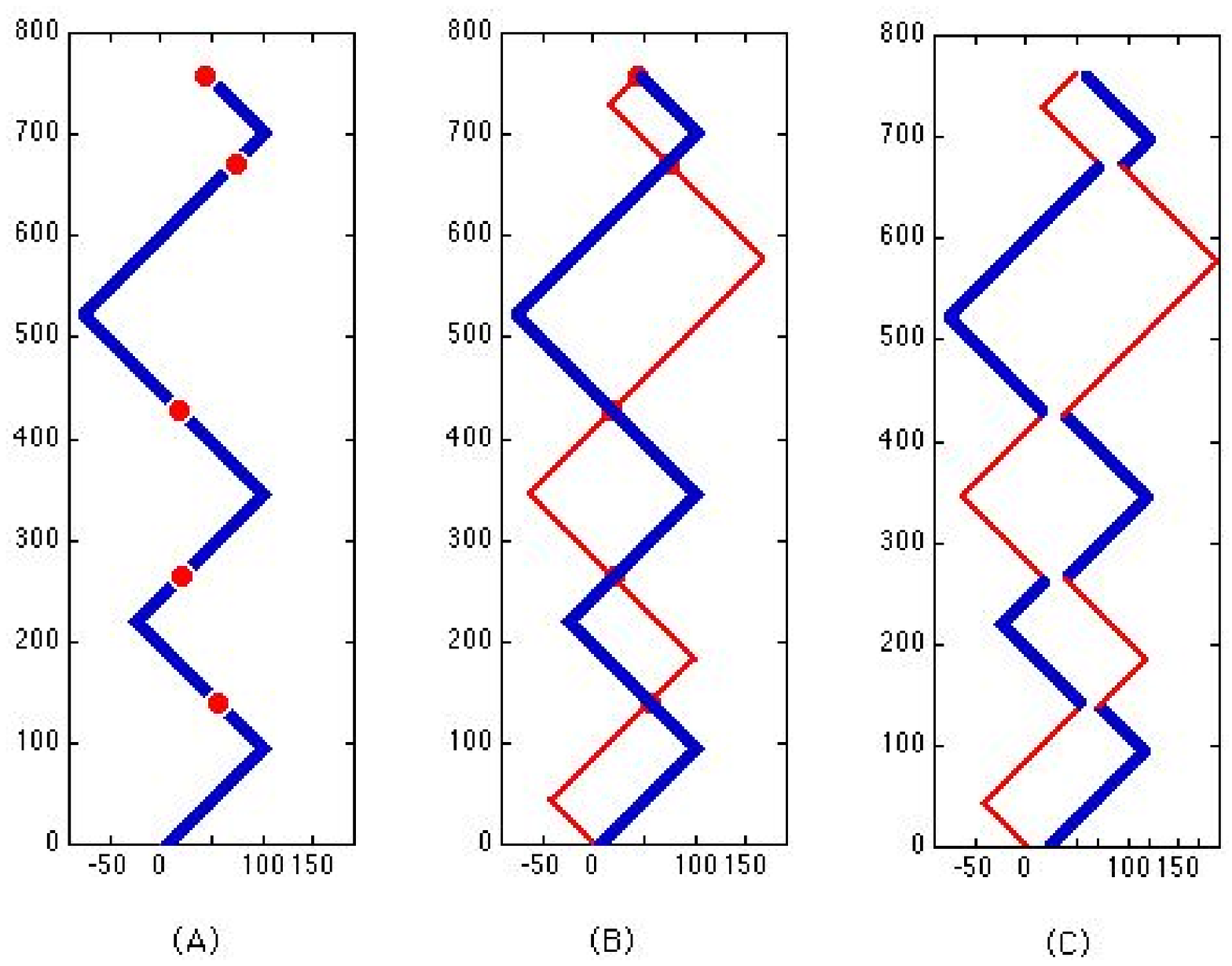}
\caption{Forming entwined paths in space-time. $z$ is horizontal, $t$ is
vertical.\newline
A) Eve travels at constant speed but occasionally reverses direction in
response to a stochastic process. At every other indication from the
stochastic process, a marker is dropped instead of a direction change (red
disks in the fig). Eve paints her path blue as she goes.\newline
B) After some specified time $t_{R}$ ($t_{R}=700$ in Fig.), Eve stops at the
next marker. She then reverses her direction in time but not in space. She
changes from blue to red and follows the `light-cone' paths through the
markers back to the origin. \newline
C) The entwined path formed in (B) can be regarded as two osculating paths
which we call envelopes. These are separated for clarity. Max paints these
envelopes with a simple rule specifying a change of colour at every second
corner, with opposite colours on the envelopes.}
\end{figure}
%%%%%%%%%%%%%%%%%%

There are several things of interest about the entwined path generated by
Eve's walk. The first thing to notice is that the distribution of distances
between the corners in the path and neighbouring markers/crossing points is
identical to the distribution of steps between the marks on the tape.
Furthermore, because Eve returns to the origin at the end of her return
journey, we see that her walk is ergodic. She can cover all such paths
simply by repeating the process enough times, covering all possible
distributions of marks on the tapes with the appropriate frequency from the
random process. Thus the time-average contribution from Eve, in the
limit of large time, is the same as the ensemble average, for when Eve has
returned to the origin after N circuits to $t_R$ and back, we cannot
distinguish the pattern of the space-time path from that of N Eves each
traversing the lattice once.

We also note that an entwined pair generated by Eve can
be viewed as two osculating paths which start at the origin moving in
opposite directions and finally merge where Eve changes directions in
$t$. We call these two paths the left and right envelopes of the pair.

The envelopes themselves have simple properties. They always have opposite
colouring. They both change colour at every second corner of the envelope.
The distribution of lengths between corners is the same for both envelopes
and is the same as the distribution of waiting times between the marks on
the tape.

The simple properties of the envelopes allow Max, our second walker, to 
traverse and colour the paths in a different way. Max will not actually step
between each neighbouring lattice site as did Eve. We imagine Max to have
long arms and he will paint entwined paths as pairs, painting one envelope
with each hand. Max uses the same tapes as Eve but he will interpret them in
a different way.

Max reads the tape out to the end of the second mark at, say, $t_{2}$. He
notes the position of the two marks on the tape and walks to the position
where the envelopes touch at the second mark. As he does this, he paints the
right envelope path blue with his right hand and the left envelope path red
with his left hand. Max needed to see where the first two marks on the tape
were in order to get both corners on his right and left hand in the
appropriate places. At $t_{2}$ Max interchanges the paintbrushes, reads the
tape out to the fourth mark and repeats his double-handed painting. Max
continues this process until he brings the paint brushes together at or
beyond the return time $t_{R}$. At this point, Max picks up his brushes and
returns to the origin without painting the lattice. Max can cover the entire
ensemble of paths by repeating the process a large number of times however,
since he always returns to the origin without painting, Max's technique is
not continuously Ergodic. Note that Eve always painted wherever she moved,
including on the return path to the origin. Eve can fully paint any of Max's
paths without having to remove her brushes from the lattice. Note also that
a single tape codes for an entwined \emph{pair} for both Eve and Max. Eve
traverses the pair in a forward and reversed path (Fig. 2a); Max `traverses'
the pair by painting the two envelopes in parallel.

Although Max's technique is not Ergodic, it can be made Markovian. That is,
if we allow Max enough arms, and the ability to paint many pairs in
parallel, he can paint the entire ensemble of paths in a single pass from
the origin to $t_{R}$, without ever requiring information beyond the current
symbols on an ensemble of tapes, and the current state of each `arm'. Thus,
he paints at step $n$ according to his state at the previous step. He never
has to reverse his direction in $t$, and he never has to read more than one
symbol per envelope path at a time. He doesn't need to read out to the
second mark on each tape before he starts painting.

To see how he does this, suppose that Max takes one of Eve's Tapes, $T_p $
say, and regards it as instructions for his right hand only. He interprets
it in the following way. Each mark will now correspond to a direction change
and every other mark will also be interpreted as a colour change, so he only
needs to read the tape one symbol at a time. Although this interpretation of
the tape by Max will not generate the same right envelope for him as it
would for Eve, the difference is a simple, unique permutation of the symbols
on the tape. That is, Max's right envelope for $T_p$ will correspond to
Eve's right envelope for another unique tape $T_{p^{\prime}}$ and vice
versa. If Max is going to paint all distinct tapes in the ensemble
simultaneously, he can use his new interpretation of Eve's tapes to do this,
since the mapping between $T_p$ and $T_{p^{\prime}}$ is invertible. Thus,
although Eve's traversal of entwined paths is not Markovian, the ensemble
average generated by her traversing the full ensemble \emph{is} Markovian,
and we may write a difference equation for it just by noting how Max
interprets and paints trajectories. This is true for both envelopes.

Now we have been thinking of $t$ as a spatial coordinate, but since Max can
generate the ensemble of paths via a Markov process which just steps forward
in $t$, we can think of $t$ as a macroscopic time. In this case the Red/Blue
colouring of Eve's path, which indicated the direction in $t$ of the
traversal, now indicates particle-antiparticle status. For example, Blue
indicates particle and Red indicates antiparticle. If we associate a plus 1
with blue we have to associate a minus one with red. This number associated
with colour we shall call charge, but it is not to be confused with
electromagnetic charge. Our charge is a classical concept which is
associated with the (discrete) continuity of the trajectory. Since our
entwined pairs all return to the origin, we see them as either pairs of
points at fixed t, or as two points superimposed, or no points at all at
fixed t. Particle number is not conserved in $t$ but charge is, if we allow
particle and antiparticle opposite charge.

How does charge behave in ensembles of entwined pairs? This is easy to
calculate if we consider ensembles of paths by their envelopes, generated by
Max's Markovian method.

Consider the left envelope path in Fig.1C. Note that the rule for its
generation is very simple. Starting at the origin, the particle proceeds in
the $-z$ direction until a mark on the tape indicates a direction change. At
the first (\emph{right}) turn, the particle just changes direction but not
colour. At the second (\emph{left}) turn the path also changes colour, and
the charge changes sign. This process is repeated. Each right corner
maintains the colour, each left turn changes the colour and the sign of the
charge. The reader familiar with the Feynman chessboard model will recognize
this rule as a version of Feynman's corner rule\cite{gord92chess,noyes}. In
this context the rule is dictated by the geometry of entwined paths. Left
turns in the left envelope are actually crossing points of the
particle-antiparticle pair, and the origin of the sign change is physical.

If we now let $\phi _{n}^{1}(z)$ be the ensemble charge density from left
envelope links parallel to the left light-cone at step $n$ and $\phi
_{n}^{2}(z)$ be the ensemble charge density from left envelope links
parallel to the right light-cone, we can write:

\begin{eqnarray}
\phi _{n}^{1}(z) &=&(1-a\Delta t)\phi _{n-1}^{1}(z+c\Delta t)-a\Delta t\phi
_{n-1}^{2}(z-c\Delta t)  \label{dirdiff1} \\
\phi _{n}^{2}(z) &=&(1-a\Delta t)\phi _{n-1}^{2}(z-c\Delta t)+a\Delta t\phi
_{n-1}^{1}(z+c\Delta t)  \nonumber
\end{eqnarray}
That is, regarding Fig. 3, most paths maintain their direction and colour as
they pass through a lattice site. The proportion which do this is $%
(1-a\Delta t)$. However a proportion $a\Delta t$ change direction at the
site. When they scatter from the right light cone they change charge on
scattering, so they \emph{decrease} the net charge in the new direction in
proportion to the density in the old direction. However when they scatter
from the left light cone they maintain their charge on scattering, so they 
\emph{increase} the net charge in the new direction in proportion to the
density in the old direction.

The right envelope is similar, except here it is the right turns which
change charge. If we let $\phi _{n}^{3}(z)$ and $\phi _{n}^{4}(z)$ be the
right envelope charges parallel to respectively left light cones and right
light cones we have 
\begin{eqnarray}
\phi _{n}^{3}(z) &=&(1-a\Delta t)\phi _{n-1}^{3}(z+c\Delta t)+a\Delta t\phi
_{n-1}^{4}(z-c\Delta t)  \nonumber \\
\phi _{n}^{4}(z) &=&(1-a\Delta t)\phi _{n-1}^{4}(z-c\Delta t)-a\Delta t\phi
_{n-1}^{3}(z+c\Delta t).  \label{dirdiff2}
\end{eqnarray}

Note that the change in signs of the scattering terms in the above equations
are a direct result of the geometry of entwined pairs. It is Eve's
insistence on entwining forward and reversed paths that allows Max to use
his simple colouring rule of a change in colour after every second corner.
This rule in turn forces the ensemble to alternate the signs of the
scattering terms, since it is a detailed feature of every path in the
ensemble.

Equations (\ref{dirdiff1}-\ref{dirdiff2}) constitute a set of coupled
difference equations in the four densities $\phi^{1}-\phi^{4}$. Although
their derivation is straightforward, their consequences as a description of
a classical ergodic stochastic process are potentially far-reaching, since
these are representations of the discrete Dirac equations -this will be
demonstrated in section IV. Furthermore, the above arguments discuss only
ensemble averages. It is possible that the above equations are correct for
the ensemble average, but that the underlying stochastic process gives
rise to such large fluctuations that the ensemble average survives only in
the event of a uniformly covered ensemble. In this case, normal stochastic
fluctuations would swamp the signal and the system would not exhibit the
above equations except under the rare circumstance of an almost perfectly
uniform coverage of the ensemble. In other words, if we watch Eve
sequentially follow a large number of randomly chosen tapes, adding and
subtracting her contributions to the sample average, would this sample
average
$\tilde{\phi}$ converge to the ensemble average predicted in
(\ref{dirdiff1}-\ref{dirdiff2})? In the following section we test this
question by stochastic simulations of Eve's generation of entwined paths. 
%%%%%%%%%%%%%%%
\begin{figure}[tbp]
\includegraphics[scale = .5]{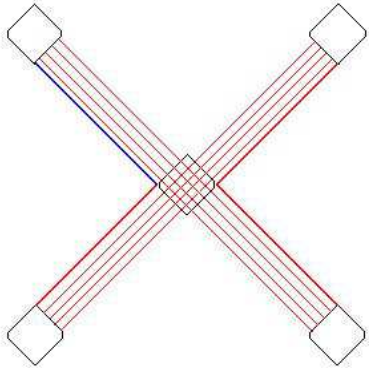}% needs
%\usepackage{graphics} in
%preamble
\caption{Left envelope scattering. Most paths do not scatter but those which
do scatter behave differently depending on which direction they are coming
from. Paths scattering from right-moving to left-moving change colour
(charge) when they scatter. Paths scattering from left-moving to right-
moving do not change colour when they scatter. }
\end{figure}
%%%%%%%%%%%%%%%%%%

\section{Numerical Checks}

In the previous section we discussed entwined paths from the perspective of
two walkers, Eve and Max. Eve draws a single entwined path by traversing a
pair through a forward passage followed by the entwined reverse passage.
Since she returns to the origin after each pair traversal, she can
sequentially cover all members of the ensemble of paths. Each time she
traverses the lattice, she records her passage by adding and subtracting
ones as appropriate to the direction she is moving in $t$ to her record of
of path visits. She thus creates a space-time field which we shall denote by 
$\tilde{\phi}$. For example, before she starts, all the $\tilde{\phi}$ are $0
$ for all $z$ with $t>0$. At her first step she arrives at $(\delta
,\epsilon )$ and adds a $1$ to $\tilde{\phi}^{4}(\delta ,\epsilon )$. If
her tape gives her a $U$ at this point, her second step is to $(2\delta
,2\epsilon )$ where she will add a $1$ to $\tilde{\phi}^{4}(2\delta
,2\epsilon )$. If she receives an $M$ at the second step she would instead
move to $(0,2\epsilon )$ where she would add a $1$ to
$\tilde{\phi}^{3}(0,2\epsilon )$. She repeats this process until the return
time where she traverses the return path, adding -$1$'s to the appropriate
$\tilde{\phi}$ on the way. If Eve were to perform this task so as to cover
all distinct paths exactly once, the  arguments of the previous section
imply that the
$\tilde{\phi}$ will satisfy the difference equations
(\ref{dirdiff1}-\ref{dirdiff2}) exactly, since these difference equations
represent the ensemble averages that Max would generate if he painted all
the paths simultaneously with his Markovian algorithm. As Eve sequentially
paints path after path, a question that needs testing is whether the
difference equations (\ref{dirdiff1}-\ref{dirdiff2}) emerge as approximate
descriptions of the sample averages $\tilde{\phi}$ generated by Eve. With
no extra controls over the stochastic process the probabilities $%
\alpha =a\Delta t$ and $\beta=1-\alpha $ as actual frequencies will
necessarily fluctuate over the sample. It is important to check that the
fluctuations do not destroy the signal, and that Eve's sequentially
generated sample does approximately satisfy the difference equations. To
check this we note that on the lattice the $\tilde{\phi}^{i}$ at any
time-step are vectors indexed by the value of
$ z$. At time step $m$ each vector $\tilde{\phi}$ has roughly $m+1$ non-zero
elements. If $||.||$ is the Euclidean norm we can then test the relative
measure of error in each separate difference equation in (\ref{dirdiff1}-\ref%
{dirdiff2}) \emph{for a sample average generated by Eve.} For example to
test equation (\ref{dirdiff1}) for a sample average $\tilde{\phi}^{1}$ we
consider the error $E_{n}^{1}$ after Eve has traversed the lattice $n$ times:

%%%%%%%%%%%%%%%
\begin{figure}[tbp]
\includegraphics[scale = .8]{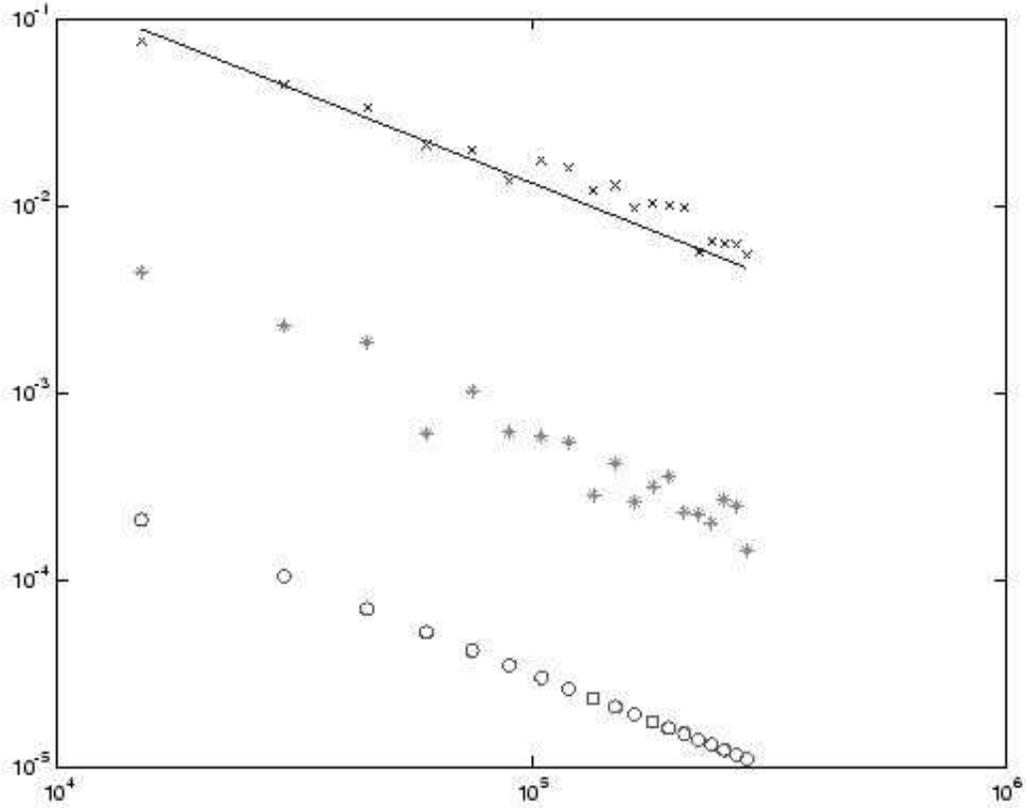}% needs
%\usepackage{graphics} in
%preamble
\caption{A numerical test of equation \ref{dirdiff2} using \ref{error}. The
numerical error is plotted vs. the number of entwined paths in the run. The
upper sequence is at $t_R$ which is 16 time steps for this run. The solid
line is a reference line which decreases as $1/n$, suggesting that for this
stochastic process the error goes down as $1/n$. The middle sequence is at $%
t=8$ and the lowest sequence is at $t=2$. As $t$ increases the error
increases since the configuration space is larger and the $\tilde{\protect%
\phi} _{i}\;$ less well covered by the sample. The tendency to converge as $%
1/n$ appears to hold for all $t$.}
\label{matlab}
\end{figure}
%%%%%%%%%%%%%%%%%%

\begin{equation}
E_{n}^{1}=\frac{||\tilde{\phi}_{n}^{1}(z)-\left( (1-a\Delta t)\tilde{\phi}%
_{n-1}^{1}(z-c\Delta t)-a\Delta t\tilde{\phi}_{n-1}^{2}(z-c\Delta t)\right)
||}{||\tilde{\phi}_{n}^{1}(z)||}  \label{error}
\end{equation}%
Analogous expressions apply for the other sample averages. Here $a$ is the 
\emph{ensemble parameter} not the sample approximation. If we were to
replace the $\tilde{\phi}$ in (\ref{error}) by the ensemble average $\phi$
we would get zero. However $E_{1}^{1}>0$ since, after  only one traversal,
the ensemble averages cannot be met, there being only a single entwined path
through the lattice. If however $\lim_{n\to \infty} E_{n}^{1} =0$ then  the
stochastic process is stable and the ensemble average description is a valid
approximation of the process for large $n$.

Figure(\ref{matlab}) shows $E_{n}^{4}$ for various values of the sample size 
$n$ plotted on a log-log scale. The three groups of points represent three
values of $t$. As $t$ increases, the range of $z$ increases and the relative
coverage by the sample paths decreases, resulting in a larger error.  If Eve
randomly chooses at each lattice site whether or not to scatter we would
expect the convergence to have the usual $1/\sqrt{n}$ dependence
characteristic of random sampling without replacement. This would yield a
very slow rate of convergence. This does indeed happen, so  we have modified
the underlying stochastic process to a form of sampling without replacement.
In our approach, when Eve has to make a decision at a lattice site, she
checks her previous decisions and chooses to scatter or not scatter based on
her local scattering probability, making the choice that causes this
probability to come closer to the ensemble average if this is possible,
choosing randomly if not. This improves the convergence so that it is
asymptotically $1/n$.  The other three errors
$E_{n}^{1},E_{n}^{2},E_{n}^{3}$ show the same convergence characteristics.
 Note that the sampling technique only speeds up the
convergence. It does not affect the limit itself since it does not alter
the ensemble average for the probabilities $\alpha$ and $\beta$. Random
sampling with replacement in which the `local' probability is not
consulted and each call to the random process is independent gives rise to
figures similar to Fig. 4 except the slope of the trend lines is less,
corresponding to
$1/\sqrt{n}$ convergence.

%%%%%%%%%%%%%%%
\begin{figure}[tbp]
\includegraphics[scale = .5]{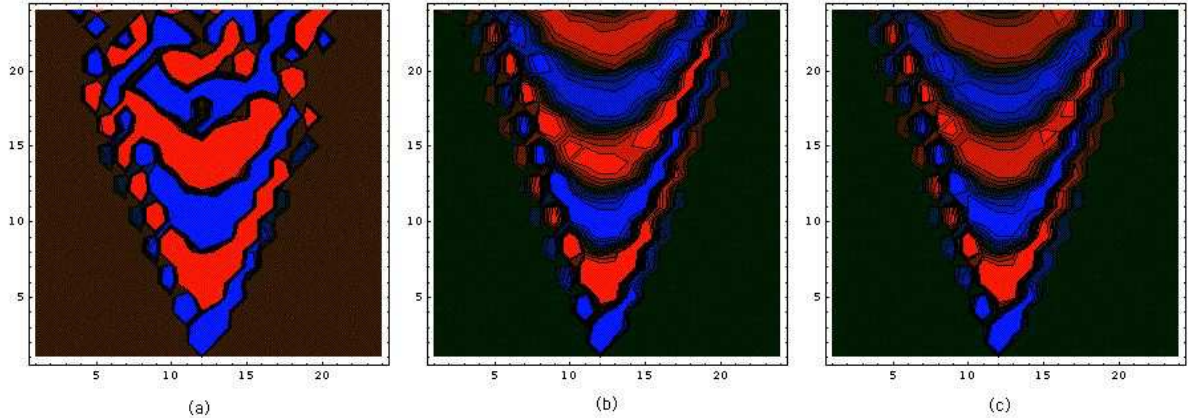}% needs
%\usepackage{graphics} in
%preamble
.
\caption{Entwined paths draw the Dirac propagator. Pictured are contour
plots of $\tilde{\protect\phi}_{3}$ for respectively $10^{3}$, $10^{5}$ and $%
10^{7}$ entwined paths. In the simulation the return time is 24 lattice
steps and the probability of scattering is $\protect\alpha \Delta t=1/2$.
(c) is visually indistinguishable from the ensemble average solution
obtained by solving (\ref{dirdiff2}) exactly. The sawtooth appearance of the
light-cone is an artifact of the lattice used to store the $\tilde{\protect%
\phi}$}
\label{contour}
\end{figure}
%%%%%%%%%%%%%%%%%%

Fig.(5) shows a contour plot of $\tilde{\phi_3}$ as Eve continues to
traverse entwined paths. At the resolution of the figure, part (c) is
indistinguishable from the exact solution of equations(\ref{dirdiff2}) with
the source at the origin. That is, Eve's formation of the component of the
propagator pictured in Fig. 5c is indistinguishable from Max's formation of
the ensemble average using his Markovian technique. The other components are
similar. Interestingly, whereas Max `draws' Fig. 5c sequentially from $t=0$
to $t=t_R$, one step at a time, colouring all paths in parallel and evolving
the `picture' in a form of unitary evolution, Eve assembles the picture as a
projection from a larger space. If we think of Eve as having her movement in
the $(z,t)$-plane parameterized by some variable $s$, then the sequence of
pictures in Fig. 5 represents a projection of Eve's history onto the $(z,t)$%
-plane with $s$ increasing from (a) to (c). Projection has been used before
to bypass FAC \cite{gnocsf96}

These numerical results confirm the fact that the difference equations (\ref%
{dirdiff1}-\ref{dirdiff2}) have an underlying stochastic process and are
phenomenological equations for entwined pairs. In the next section we show
that the difference equations describing the ensemble averages are discrete
versions of the Dirac equation.

\section{Comparison with Dirac}

By expanding equations (\ref{dirdiff1}-\ref{dirdiff2}) to first order we can
approximate the difference equations by a set of coupled PDE's. The
resulting equations are: 
\begin{eqnarray}
\frac{\partial \phi ^{1}}{\partial t} &=&c\frac{\partial \phi ^{1}}{\partial
z}-a\phi ^{1}-a\phi ^{2}  \nonumber \\
\frac{\partial \phi ^{2}}{\partial t} &=&-c\frac{\partial \phi ^{2}}{%
\partial z}-a\phi ^{2}+a\phi ^{1} \\
\frac{\partial \phi ^{3}}{\partial t} &=&c\frac{\partial \phi ^{3}}{\partial
z}-a\phi ^{3}+a\phi ^{4}  \nonumber \\
\frac{\partial \phi ^{4}}{\partial t} &=&-c\frac{\partial \phi ^{4}}{%
\partial z}-a\phi ^{4}-a\phi ^{3}  \nonumber
\end{eqnarray}
Removing the exponential decay by writing $u^{\mu }(z,t)=e^{at}\phi ^{\mu
}(z,t)$ the above becomes 
\begin{eqnarray}
\frac{\partial u^{1}}{\partial t} &=&c\frac{\partial u^{1}}{\partial z}%
-au^{2}  \nonumber  \label{dirac4} \\
\frac{\partial u^{2}}{\partial t} &=&-c\frac{\partial u^{2}}{\partial z}%
+au^{1}  \nonumber \\
\frac{\partial u^{3}}{\partial t} &=&c\frac{\partial u^{3}}{\partial z}%
+au^{4} \\
\frac{\partial u^{4}}{\partial t} &=&-c\frac{\partial u^{4}}{\partial z}%
-au^{3}  \nonumber
\end{eqnarray}%
The above is a representation of the Dirac equation in which all the
densities $u$ are real. This may be seen by writing the equation in matrix
form. If we write $p_{z}=-i\frac{\partial }{\partial z}$, setting $c=1$ and $%
a=m$ with $\alpha _{z}=-\left( 
\begin{array}{cc}
\sigma _{z} & 0 \\ 
0 & \sigma _{z}%
\end{array}%
\right) $ $\beta =\left( 
\begin{array}{cc}
\sigma _{y} & 0 \\ 
0 & -\sigma _{y}%
\end{array}%
\right) $ we have 
\begin{equation}
i\frac{\partial u}{\partial t}=(\alpha _{z}p_{z}+\beta m)u  \label{dirac1}
\end{equation}%
where the $\sigma$ are the Pauli matrices. Note that $\alpha $ and $\beta $
anticommute as required, and the relativistic energy-momentum relations are
obeyed if we associate the usual meanings with $E$ and $p$. It is important
to note that the rewrite of equations (\ref{dirac4}) in (\ref{dirac1}) is
only cosmetic. The use of $p$ and $i$ are conventions only; the $u$ are
still real densities, the limit of ensemble averages. They are not the
formal objects of conventional quantum mechanics.

\section{Discussion}

The Dirac equation is usually produced along the lines of Dirac's original
argument. The start of the argument is canonical quantization $p\to -i\hbar 
\frac{\partial}{\partial x}, E\to i\hbar \frac{\partial}{\partial t}$. From
there Dirac leads us through the construction of his algebra to satisfy the
requirements of the relativistic energy-momentum relations. From the
perspective of the above work, the formal step in Dirac's argument is the
first one, the FAC.

Entwined paths are essentially self-quantizing, as can be seen by the form
of (\ref{dirac1}). If $m$ is set equal to zero in (\ref{dirac1}) the
resulting equation is just two, 2-component forms of the wave equation which
are appropriate for classical particles that stay on their initial light
cones. It is the $m\beta $ term that both describes scattering and brings in
the interference effects characteristic of the Dirac equation. An inspection
of the $\beta $ matrix shows that its hermitian character arises because of
the factor of $i$ absorbed into it in the rewrite from (\ref{dirac4}). The
predecessor of $\beta$ in (\ref{dirac4}) is \emph{antihermitian} and it is
this feature that ultimately results in interference effects. On the other
hand the antihermitian form arises \emph{directly} from the entwined
geometry of the paths. It is for this reason that entwined paths make the
canonical quantization step from classical physics unnecessary. The
space-time geometry of entwined paths automatically builds in the relevant
physics. To see this, note that it is the regular crossing of the entwined
paths that produces the alternating signs of the scattering terms in (\ref%
{dirdiff1}-\ref{dirdiff2}), and ultimately the $\beta $ matrix in (\ref%
{dirac1}) .

The existence of an underlying stochastic model for the Dirac equation, as
demonstrated in this paper, allows us the opportunity to consider the Dirac
equation as a phenomenological equation describing the evolution of a
particle moving on an entwined path in space-time. We now have a stochastic
basis for the ``U process'' of quantum mechanics (that process by
which a wavefunction unitarily evolves), to use Penrose's terminology
\cite{Penrose89,Penrose95}, at least in the special case of a free
particle in one dimension. We hope exploration of the stochastic model will help to
further clarify the relationships between classical and quantum physics, and
possibly shed some light on the  process by which
wavefunctions `collapse' as a consequence of a measurement (the
`R''-process).

\begin{acknowledgments}
GNO is grateful for helpful discussions and support from J.A. Gualtieri.
This work was partially supported by the Natural Sciences and Engineering
Research Council of Canada.
\end{acknowledgments}

%\bibliography{index}

\end{document}